\documentclass[final,3p,times,twocolumn]{elsarticle}

\usepackage{amsmath,graphicx}

\begin{document}

\begin{frontmatter}

\title{Simulation of Self-Pulsing in Kerr-Nonlinear Coupled Ring Resonators}

\author{Ji{\v{r}}{\'{\i}} Petr{\'{a}}{\v{c}}ek\corref{cor1}}
\ead{petracek@fme.vutbr.cz}
\cortext[cor1]{Corresponding author. Tel.: (+420) 541 142 764; fax: (+420) 541 142 842.}
\author{Yasa Ek\c{s}io\u{g}lu}
\author{Anna Sterkhova}

\address{Institute of Physical Engineering, Faculty of Mechanical Engineering, Brno University of Technology,  Technick{\'{a}} 2, 616 69 Brno, Czech Republic}

\begin{abstract}
Nonlinear resonant structures consisting of coupled ring resonators can be modeled by difference-differential equations that take into account 
non-instan\-taneous Kerr response and the effect of loss. We present a simple and efficient numerical formalism for solution of the system and calculation of the time evolution. 
The technique is demonstrated by investigating the dynamical behavior of the coupled structure with two rings, namely, focusing on self-pulsing solutions. The influence of both, loss 
and non-instantaneous Kerr response, is also presented.
\end{abstract}


\end{frontmatter}


\section{Introduction}
\label{Introduction}
A number of interesting dynamical phenomena can occur in optical systems with nonlinear feedback. Such systems, typically nonlinear cavities, can, within a certain range of input optical power,
exhibit optical bistability (or even multistability), a phenomenon which is expected to play an important role in data processing applications. With a further increase of power, the systems may exhibit generation of optical pulses from continuous wave input (self-pulsing) and chaos. The nonlinearity is often provided by the Kerr effect, in which the intensity of propagating light alters the refractive index of the medium. 

The chaotic behavior is related to the instability that was first investigated by Ikeda \cite{Ikeda1} in a single ring resonator. Under suitable conditions, namely for the finite relaxation time of the nonlinear response, the Ikeda instability can lead to self-pulsing (SP) \cite{Ikeda2}. These predictions were experimentally confirmed with a hybrid optically bistable device \cite{gibbs}. Similar effects were found in a nonlinear Fabry-Perot resonators \cite{ogusu} and distributed feedback structures \cite{winful,parini}. The impact of the relaxation time was recently studied in Ref. \cite{armaroli}

Other nonlinear mechanisms can also induce the mentioned effects. For various types of optical microcavities, e.g., silicon microring resonators, the nonlinear effect is not only provided by the Kerr effect but also by the free carrier absorption (FCA) and free carrier dispersion (FCD) through the two-photon absorption (TPA) effect \cite{priem,xu,Malaguti,chen,Vaerenbergh,Zhang}.

It is well-known that nonlinear effects are enhanced in coupled-cavity systems \cite{melloni,Blair}. Coupled cavities with instantaneous Kerr response exhibit rich dynamics and offer more control over nonlinear switching, SP and chaos \cite{dumeige,petracek2,maes,grigoriev,petracek}. In particular, SP and chaos were observed in systems with two and three coupled microcavities \cite{maes}. For two-cavity systems, SP can be explained as a result of beating of modes and bistable switching \cite{grigoriev}. However, SP can also be related to gap solitons \cite{maes}. For long microring chains, spontaneous generation of gap solitons from cw input was studied in Ref. \cite{Posada}. 

In this paper, we focus on simulation of coupled ring resonators. Compared with the publications regarding coupled-cavity systems, we take into account non-instantaneous Kerr response. We also introduce the effect of loss which has not been considered in such systems so far.

Often, the simulation of nonlinear cavities is performed by coupled mode theory in time \cite{maes,grigoriev,armaroli,dumeige4}. The technique can treat different kinds of cavities in the same manner. However, its validity is limited to the case of weak coupling \cite{haus}. Here we describe an alternative approach that can be used for ring resonators. We generalize the map presented in Ref. \cite{Ikeda2} and obtain a system of difference-differential equations. We formulate an efficient numerical technique for the solution of the system. The paper is organized as follows: In Section \ref{tech}, we introduce the model and theoretical formulation. In Section \ref{results}, we demonstrate the technique by presenting the dynamical behavior of a system with two rings. The effect of loss as well as the effect of instantaneous and non-instantaneous Kerr response is also presented. Finally, in Section \ref{conc} we conclude the paper. 

\begin{figure*}
\centering
\includegraphics[width=12.3cm]{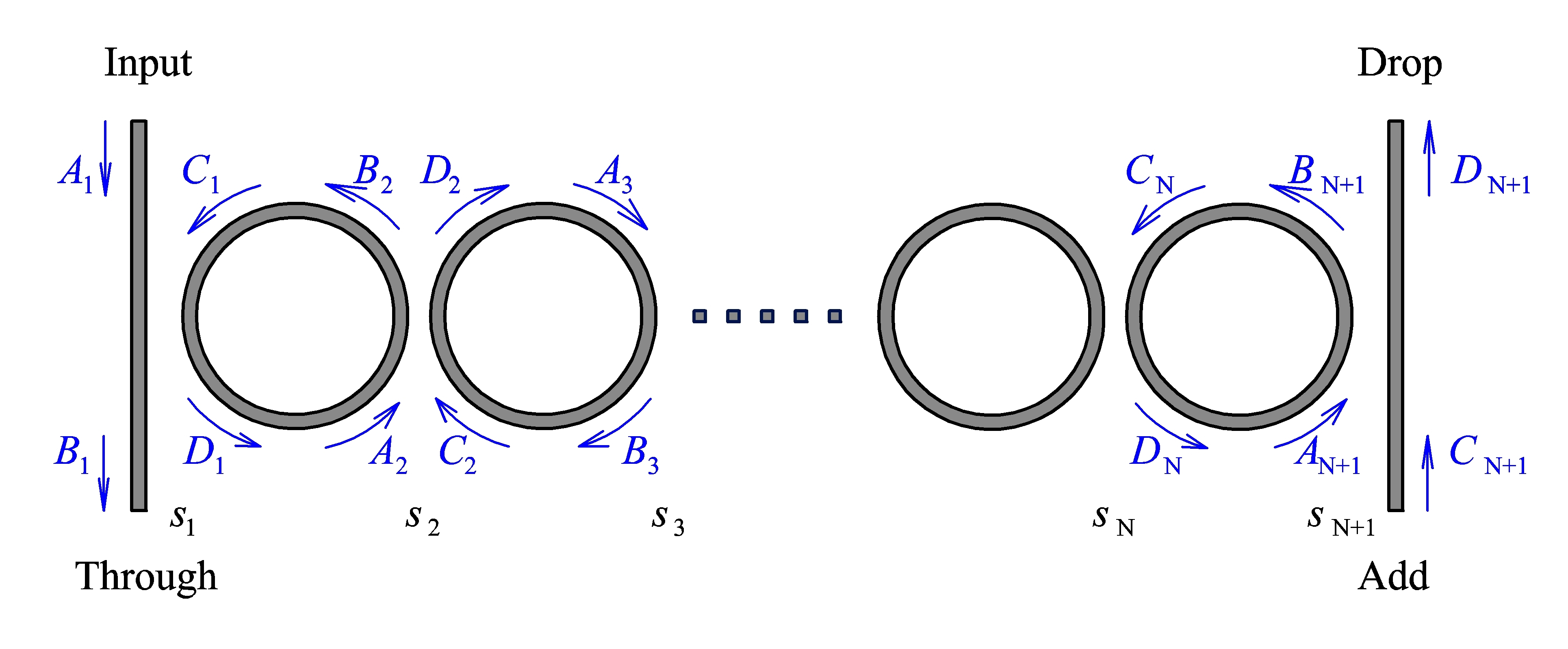}
\caption{Coupled ring resonators with four ports labeled. Each coupler is described by the parameter $s_j$. $A_j$, $B_j$, $C_j$ and $D_j$ represent mode amplitudes. Arrows indicate propagation of the modes.}
\label{fig:adf}
\end{figure*}

\section{Model and Theoretical Formulation}
\label{tech}
Consider a structure consisting of $N$ coupled ring resonators side coupled with two waveguides as shown in Fig. 1. All the resonators are identical and made from the same single mode waveguide as the two waveguides. $A_j$, $B_j$, $C_j$ and $D_j$ represent the time-dependent (slowly-varying) mode amplitudes at different positions in the rings or waveguides. The amplitudes are scaled to dimensionless form, as will be explained below. The presented model is valid for unidirectional propagation of modes, i.e. the structure is excited at the input and/or add port. In the subsequent numerical calculations (Section 3), however, we will always suppose excitation only at the input port, thus $C_{N+1}=0$. 

Similarly as in \cite{Yariv2002,Poon2003}, we assume that the coupling is lossless and localized at a single point. Then, by using the notation in Fig. 1, the interaction in each coupler (in time $t$) is given by
\begin{equation}
\ B_j(t)= r_j A_j(t)+is_j C_j(t),
\label{eq:ALP_dim_1}
\end{equation}
\begin{equation}
\ D_j(t)= is_j A_j(t)+r_j C_j(t).
\label{eq:ALP_dim_2}
\end{equation}
Here, $is_j$ and $r_j=\sqrt{1-s_{j}^2}$, ($j=1, 2, \dots , N+1$), are respectively the coupling and transmission coefficients which describe each coupler \cite{Yariv2002,Poon2003}.

The structure exhibits Kerr nonlinearity, i.e., in the stationary state, the nonlinear change of the effective mode index at a certain position in the ring (or waveguide) is given by the relation of type 
$n_2 |A_j^\prime|^2$, where $n_2$ is the effective nonlinear Kerr-index (it is assumed to be constant in all rings/waveguides) and $A_j^\prime$ is the physical amplitude of the mode at this position. However, instead of $A_j^\prime$ we use the dimensionless amplitude $A_j$ defined by the relation $A_j=\left(2\pi n_2 L_{\rm{eff}}/\lambda\right)^{1/2} A_j^\prime$,  where $L_{\rm{eff}}=\left[1-\exp\left(-\alpha L\right)\right]/\alpha$ is the effective length, $\alpha$ is the waveguide loss coefficient (it includes all possible linear loss mechanisms, such as material absorption or scattering), $L$ is the half of the ring circumference, and $\lambda$ is the wavelength of the light in vacuum. The same scaling applies for the amplitudes $B_j$, $C_j$ and $D_j$. 
In accordance with these definitions, we define powers at the input, through, and drop ports by the relations $P_{\rm{in}}= (L/L_{\rm{eff}})\;|A_1|^{2}$,  $P_{\rm{t}}= (L/L_{\rm{eff}})\;|B_1|^{2}$, and 
$P_{\rm{d}}= (L/L_{\rm{eff}})\;|D_{N+1}|^{2}$, respectively. In this way, the powers are directly related to the physical amplitudes and can also be used as a measure of nonlinearity strength. 

Optical pulse propagation in a nonlinear dispersive media is well-described by Maxwell's equations \cite{Agrawal,Boyd} taking into consideration nonlinear polarization and applying the slowly-varying-envelope approximation. In the presence of Kerr-nonlinearity and waveguide loss, the relations between mode amplitudes in Fig. 1 are:
\begin{equation}
\ C_j(t)= B_{j+1}(t-\tau) \; \exp\left[-\frac{\alpha L}{2}+i\phi+i \beta_{j+1}(t-\tau)\right],
\label{eq:ALP_pm_1}
\end{equation}
\begin{equation}
\ A_{j+1}(t)= D_j(t-\tau) \; \exp\left[-\frac{\alpha L}{2}+i\phi+i \delta_{j}(t-\tau)\right].
\label{eq:ALP_pm_2}
\end{equation}
Here, $1\le j \le N$, $\tau=n_{\rm{g}} L/c$ is the group delay corresponding to propagation of the pulse over distance $L$, $n_{\rm{g}}$ is the mode group index and $c$ is the velocity of light. Note that, the free spectral range, FSR, is related with the group delay by the relation $\tau\,{\rm{FSR}}=1/2$. $\phi=2 \pi n_{\rm{eff}} L/\lambda$ is the linear phase shift acquired over distance $L$, $n_{\rm{eff}}$ is the linear effective mode index. The shift can be expressed as $\phi = \pi\left(m+\Delta f/{\rm{FSR}}\right)$, where $m$ is an arbitrary positive integer and $\Delta f$ is the frequency detuning from resonance. In the following analysis, we will always assume even $m$ (adaption of the formulation for 
odd $m$ is obvious) and thus $\phi$ in Eqs. (\ref{eq:ALP_pm_1}) and (\ref{eq:ALP_pm_2}) can be replaced by $\pi\,\Delta f/{\rm{FSR}}$. 

Nonlinear phase shifts $\beta_{j}$ and $\delta_{j}$ are given 
by the response of the medium. Here, we assume the Debye relaxation equations \cite{Ikeda1,Ikeda2,Moloney} 
\begin{equation}
\ T_{\rm{R}}\frac{d \beta_{j}(t)}{dt}+\beta_{j}(t)= |B_{j}(t)|^{2},
\label{eq:ALP_delay_1}
\end{equation}
\begin{equation}
\ T_{\rm{R}}\frac{d \delta_{j}(t)}{dt}+\delta_{j}(t)= |D_{j}(t)|^{2},
\label{eq:ALP_delay_2}
\end{equation}\\
where $T_{\rm{R}}$ is the medium relaxation time. 

The above system of the difference-differential equations (\ref{eq:ALP_dim_1}-\ref{eq:ALP_delay_2}), which is a generalization of the Ikeda equations for single ring \cite{Ikeda2,gibbs}, fully describes the time evolution of the amplitudes $A_j$, $B_j$, $C_j$, $D_j$ and nonlinear phase shifts $\beta_{j}$, $\delta_{j}$ from given initial conditions. 

The system can be readily solved in the approximation of instantaneous response. In this case, we consider the limit $T_{\rm{R}}<<\tau$ and assume the solutions of Eqs. (\ref{eq:ALP_delay_1}) and (\ref{eq:ALP_delay_2}) in the form $\beta_{j}(t)= |B_{j}(t)|^{2}$ and $\delta_{j}(t)= |D_{j}(t)|^{2}$. Consequently, the whole system is reduced to a difference equations which appear, e.g., in \cite{Posada}.
 
For obtaining of the steady-state solutions (time independent solutions) of Eqs. (\ref{eq:ALP_dim_1}-\ref{eq:ALP_delay_2}), we assume no signal at the add port, $C_{N+1}=0$, and arbitrarily choose the amplitude at the drop port $D_{N+1}$.  Then, the other amplitudes are calculated step by step with using Eqs. (\ref{eq:ALP_dim_1}-\ref{eq:ALP_pm_2}), finally the amplitudes $A_{1}$ at the input and $B_{1}$ at the through port are found. Note that in the steady state, solutions of Eqs. (\ref{eq:ALP_delay_1}) and (\ref{eq:ALP_delay_2}) are formally the same as in the approximation of instantaneous response.

Stability of the steady-state solutions was investigated by using the linear stability analysis. To this aim, we assumed the approximation of instantaneous response and calculated the eigenvalues of the Jacobian. The given solution is stable if and only if the absolute value of any eigenvalue is less than 1 \cite{gibbs}. 

In the general case of non-instantaneous response, we need an efficient technique for the numerical integration of 
Eqs. (\ref{eq:ALP_delay_1}) and (\ref{eq:ALP_delay_2}). To this aim, we write 
solution of Eq. (\ref{eq:ALP_delay_1}) in the form
\begin{equation}
\begin{split}
\beta_{j}(t) &= |B_{j}(t)|^{2} + \exp\left(-\frac{t}{T_{\rm{R}}}\right)\left[\beta_{j}(0)-|B_{j}(0)|^{2}\right] + \\
&-\int\limits_0^t\exp\left(\frac{t'-t}{T_{\rm{R}}}\right) \frac{d|B_{j}(t')|^{2}}{dt'} dt'.
\label{eq:sol_delay1}
\end{split} 
\end{equation}
For numerical calculation, we discretize the variable $t$ with the step $\Delta t$ and evaluate the integral on the RHS of Eq. (\ref{eq:sol_delay1}) by the midpoint rule. 
Then we obtain the following approximate solution of Eq. (\ref{eq:ALP_delay_1})
\begin{equation}
\begin{split}
\beta_{j}(t+\Delta t) & \approx |B_{j}(t+\Delta t)|^{2} + \\
& +\exp\left(-\frac{\Delta t} {T_{\rm{R}}}\right)\left[\beta_{j}(t)-|B_{j}(t)|^{2}\right] + \\
& -\exp\left(-\frac{\Delta t}{2T_{\rm{R}}}\right) \left[|B_{j}(t+\Delta t)|^{2} - |B_{j}(t)|^{2}\right],
\label{eq:appsol_delay1}
\end{split} 
\end{equation}
which is substituted into Eq. (\ref{eq:ALP_pm_1}). Analogically we find a solution of Eq. (\ref{eq:ALP_delay_2}) and substitute into Eq. (\ref{eq:ALP_pm_2}). 
With these considerations the calculation of the time evolution is straightforward.

\section{Numerical results}
\label{results}
The above described technique can simulate a device with an arbitrary number of rings. However, with increasing number of rings, the structures exhibit more complicated behavior. Therefore, with the aim of obtaining tractable results and keeping the number of structural parameters to a minimum, we present simulation of a simple structure with two rings. 

\begin{figure}[ht]
\includegraphics[width=7.7cm]{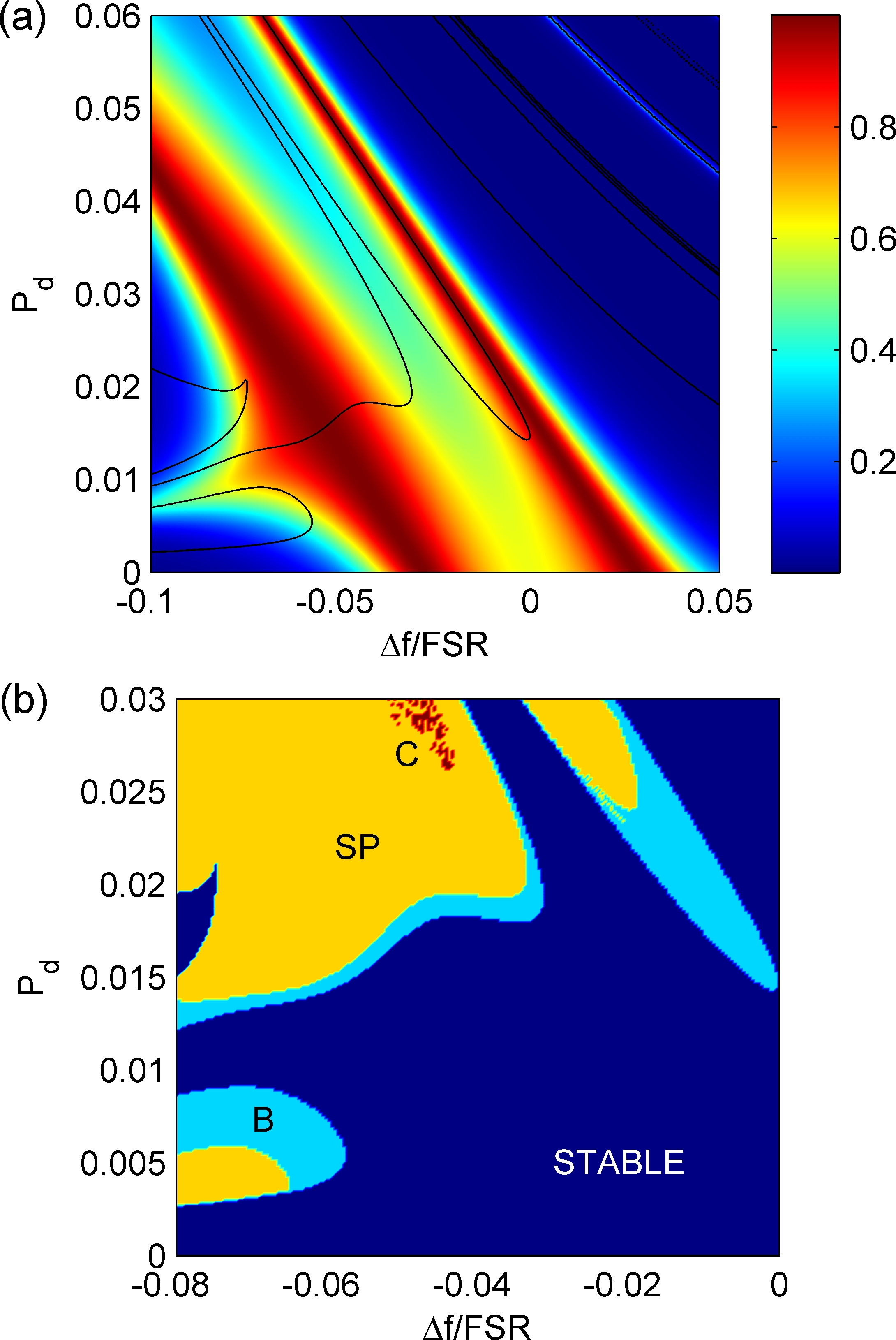}
\caption{(a) Steady-state solutions for the device with two rings. The color map indicates the relative power $P_{\rm{d}}/P_{\rm{in}}$ at the drop port as a function of power $P_{\rm{d}}$ at the drop port and relative detuning $\Delta f/\rm{FSR}$. Solid lines show boundaries between stable and unstable states. (b) Classification of the states. Unstable states are labeled as bistable/multistable (B), self-pulsing (SP), and chaos (C). 
The structural parameters are $s_{1}=s_{3}=0.42$, $s_{2}=0.2$, $\tau/{T_{\rm{R}}}=2$, $\alpha=0$; the parameter $\tau/{T_{\rm{R}}}$ is not used in calculation of (a). Note that the axes in (a) and (b) have different scales.}
\label{fig:map}
\end{figure}

Fig. \ref{fig:map} shows steady-state solutions and their classification for the selected device. Generally, it is well-known, that structures with coupled cavities can be optimized for obtaining the desired (e.g. flat-top) linear response; for structures with identical rings such optimization is done by suitable selection of coupling parameters $s_j$ \cite{little}. However, the systems optimized for flat-top response are not suitable for observing bistability, SP, and chaos \cite{Blair,maes}. Therefore, 
we select the parameters $s_j$ in such a way, that the device exhibits two resonance peaks ($P_{\rm{d}}\approx P_{\rm{in}}$), which are 
clearly seen in Fig. \ref{fig:map}(a). The behavior is similar to those observed in Ref. \cite{maes}. The peaks shift towards the negative values of the detuning as we increase nonlinearity strength (the power at the drop port $P_{\rm{d}}$). The slope of the resonance regions is (approximately) equal to $-1$, which can be qualitatively explained by the linear increase of the effective mode index $n_{\rm{eff}}$ with $P_{\rm{d}}$.

We evolve unstable solutions in time until we reach either stable, SP or chaotic solutions while keeping the input power $P_{\rm{in}}$  constant, see Fig. \ref{fig:map}(b). The stable solutions belong to bistable (or multistable) regions. In order to distinguish between SP (periodic) and chaotic solutions, we calculated autocorrelation of the signal at the drop port $D_3(t)$. In this way, we were also able to determine a period of the self-pulsations. Here, and in all the following examples, the calculation of the time evolution with non-instantaneous response was performed with the step $\Delta t=\tau / 10$ and we also verified that this value has negligible influence on the presented results.

The most interesting region for SP is found for negative detuning, below $\Delta f/\rm{FSR}\approx -0.035$, and nonlinearity levels above $P_{\rm{d}}\approx 0.02$. (SP states around  $P_{\rm{d}}\approx 0.005$ are not attractive because $P_{\rm{d}}/P_{\rm{in}}$ is small.) Note that the region extends next to the place where the resonance leaks into the gap.  For the detuning below $\Delta f/\rm{FSR}\approx -0.057$, the structure exhibits bistability (for nonlinearity levels $P_{\rm{d}}\approx 0.007$) before reaching SP. Contrary to this, above $\Delta f/\rm{FSR}\approx -0.057$, SP can occur without undergoing a bistable state. Chaotic states are found inside the SP region, near $P_{\rm{d}}\approx 0.03$ and $\Delta f/\rm{FSR}\approx -0.047$.

\begin{figure}[ht]
\includegraphics[width=7.7cm]{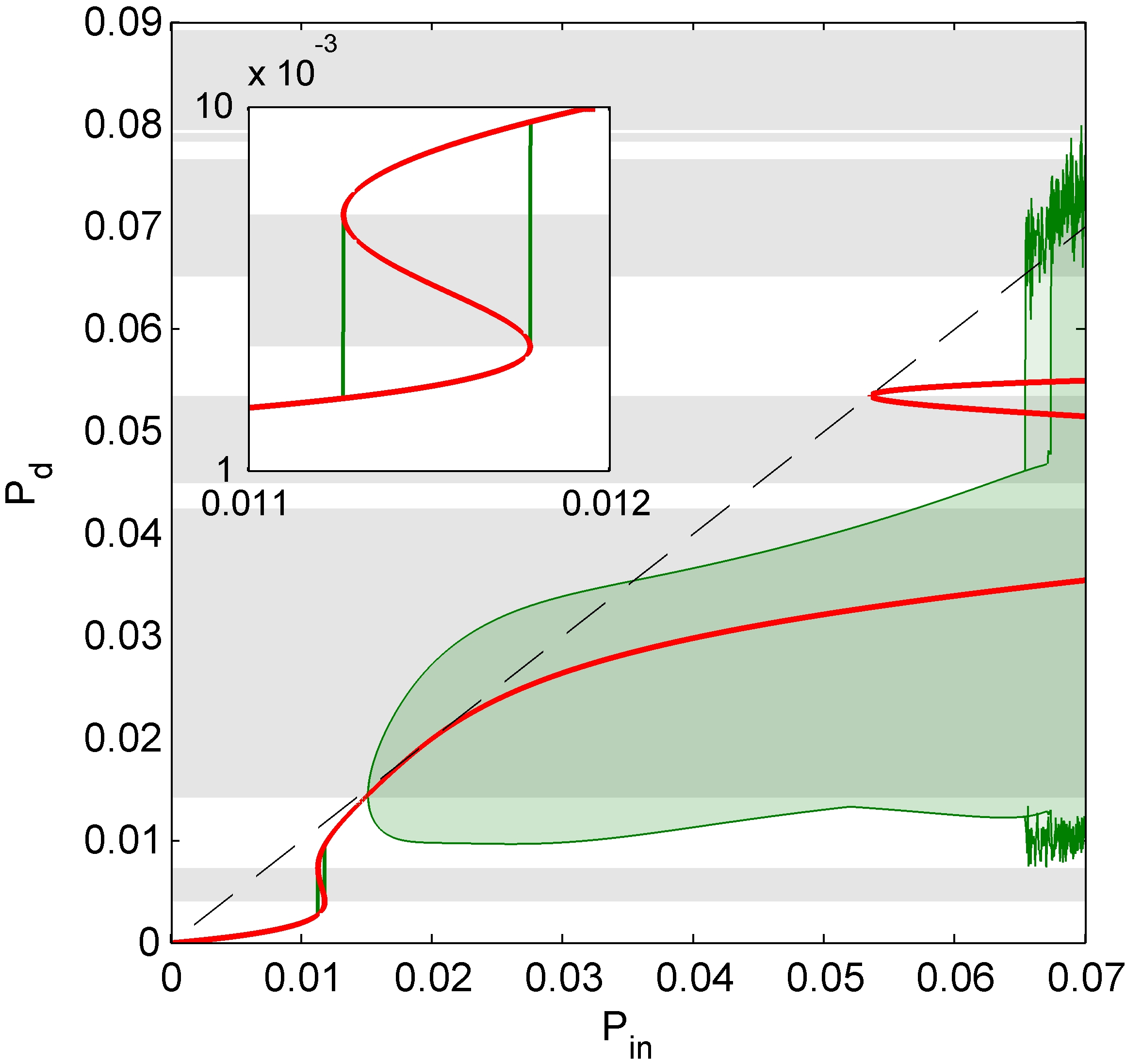}
\caption{Power at the drop port $P_{\rm{d}}$ vs. input power $P_{\rm{in}}$ for $\Delta f/\rm{FSR}=-0.06$; other structural parameters are as in Fig. 2. Green curve (area) labels values of $P_{\rm{d}}$ calculated for adiabatic increase and decrease of $P_{\rm{in}}$ (from 0 to 0.07 and back). Superimposed red curve shows steady-state solutions. Horizontal gray shaded regions mark intervals of  $P_{\rm{d}}$ in which the steady-state solutions are unstable. Dashed line denotes 100\% power transfer into the drop port, i.e., $P_{\rm{d}}=P_{\rm{in}}$. Bistable region is magnified in the inset.}
\label{fig:bifurcation}
\end{figure}

In order to examine the described behavior more closely, we consider $\Delta f/\rm{FSR}$, which belongs to the mentioned bistable region, and present a bifurcation diagram in Fig. \ref{fig:bifurcation}. [In our calculation, we, step by step, slightly increased  (or decreased) $P_{\rm{in}}$, fixed this value and searched for the corresponding stable, SP or chaotic state. As a initial condition, we used the state found for the previous value of $P_{\rm{in}}$. It should be noted that other solutions can be found for different ways of changing $P_{\rm{in}}$.] As expected, we observe a bistable region followed by a Hopf bifurcation at $P_{\rm{in}}\approx 0.016$, which is the birth of a limit cycle from the equilibrium \cite{Strogatz,Kocak}, and suggests periodic oscillations.  
SP states are observed above the Hopf bifurcation and bifurcate into chaotic states at $P_{\rm{in}} \approx 0.067$. Interestingly, there is a narrow region about $P_{\rm{in}} \approx 0.066$ 
in which we observe either SP or chaos for increasing or decreasing of $P_{\rm{in}}$, respectively. 
Note also that the unstable steady-states (gray shaded regions), which were identified by linear stability analysis with the approximation of instantaneous response, precisely correspond to the results of the exact calculation (green curve/area).

The behavior seen in Fig. \ref{fig:bifurcation} is characteristic for the given detuning. For example, as discussed before, the bistable region vanishes with increasing $\Delta f/\rm{FSR}$. On the other hand, decreasing $\Delta f/\rm{FSR}$ leads to a wider bistable region and finally to overlapping of SP and bistable regions. We found various possible scenarios which are analogical to those observed in  the case of a single cavity \cite{armaroli}.

\begin{figure}[ht]
\includegraphics[width=7.7cm]{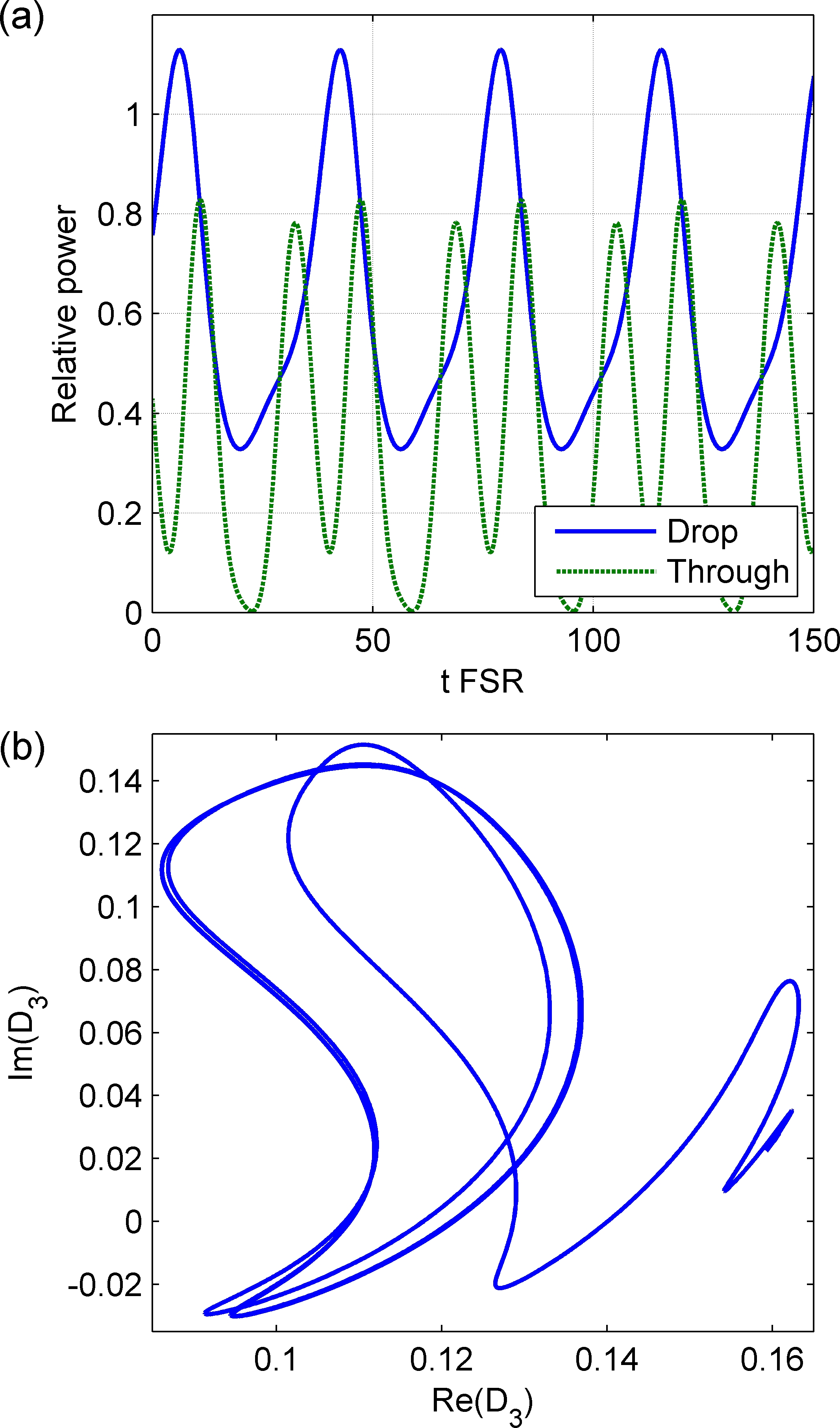}
\caption{Self-pulsing solution for the input power $P_{\rm{in}}=0.03$; other structural parameters are as in Fig. 3. (a) Relative powers $P_{\rm{d}}/P_{\rm{in}}$ and $P_{\rm{t}}/P_{\rm{in}}$ at the drop and through ports, respectively, vs. normalized time $t\,\rm{FSR}$. (b) Phase portrait of the amplitude $D_3$ at the drop port.}
\label{fig:limitcycle}
\end{figure}
\begin{figure}[ht]
\includegraphics[width=7.7cm]{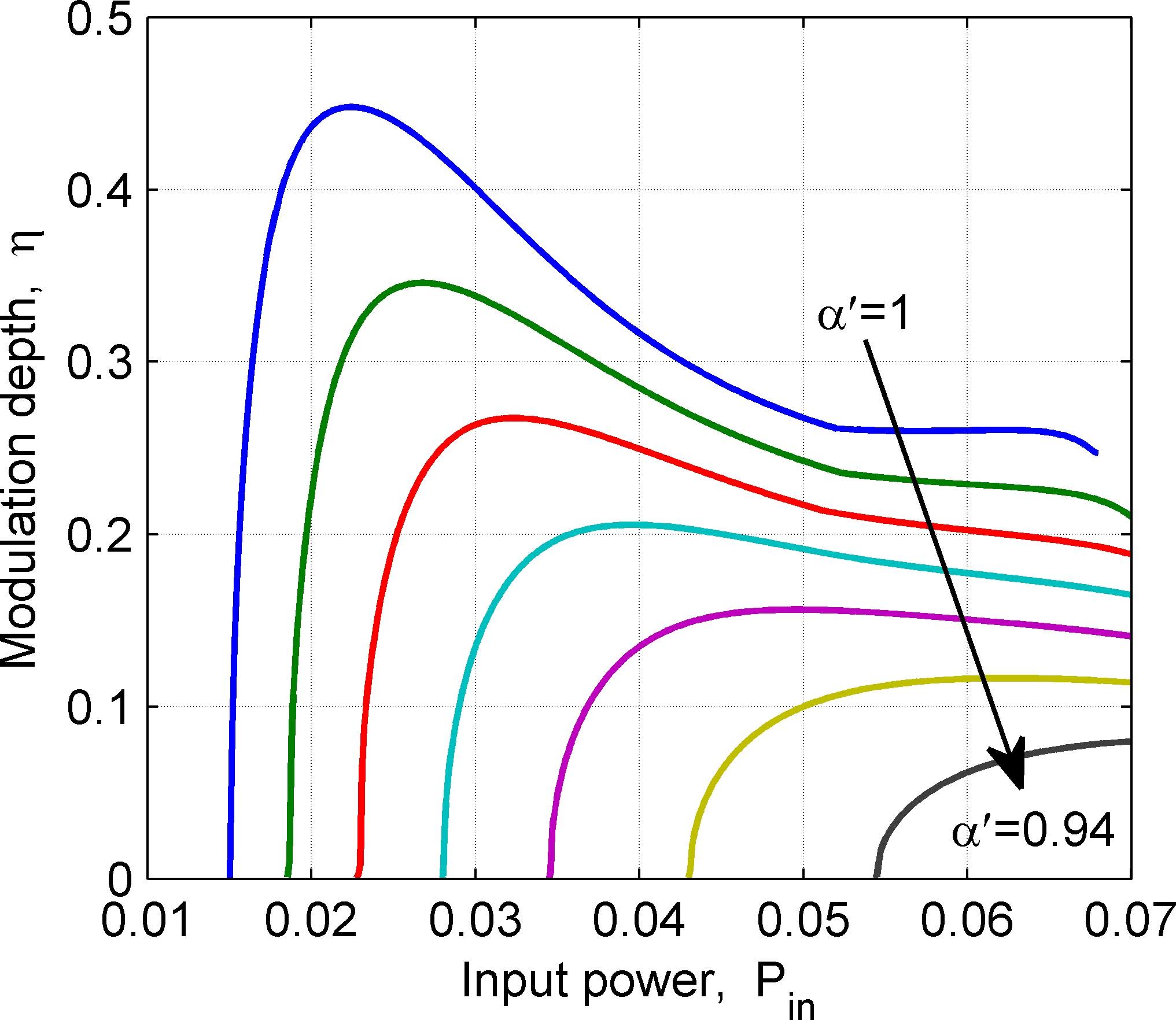}
\caption{The modulation depth $\eta$ vs. input power $P_{\rm{in}}$ for seven different values of the loss coefficient $\alpha^\prime$ ($1$, $0.99$, $0.98$, $0.97$, $0.96$, $0.95$, and $0.94$); other structural parameters are as in Fig. 3. The arrow indicates the trend of increasing loss (decreasing $\alpha^\prime$). The dependencies were calculated for the adiabatic increase of $P_{\rm{in}}$.}
\label{fig:loss}
\end{figure}

An example of SP state is shown in Fig. \ref{fig:limitcycle}(a). Obviously, the period, amplitude (cf. Fig. \ref{fig:bifurcation}) and shape of pulses depend on the selected structure parameters.
However, the pulses emerging from the through port usually have more complex shapes than the pulses from the drop port. 
The corresponding phase portrait in Fig. \ref{fig:limitcycle}(b) demonstrates an attracting limit cycle; this calculation was obtained by evolving of the steady state found for the given $P_{\rm{in}}$ (at which $D_3=0.1604 + i0.0262$). 

The effect of losses is demonstrated in Fig. \ref{fig:loss}. To this aim, we considered SP solutions and calculated the modulation depth, defined here as 
$\eta= \left[  \max\left(P_{\rm{d}}\right) - \min\left(P_{\rm{d}}\right) \right] / \left(2P_{\rm{in}}\right)$. The loss is characterized by the parameter $\alpha^\prime=\exp(-\alpha L)$.
(Note that $\eta=0$ corresponds to stable states. For chaotic states, observed for lossless case $\alpha^\prime=1$, $\eta$ is not shown in Fig. \ref{fig:loss}.)
It is seen, that the loss significantly affects the SP behavior, namely, it decreases the modulation depth and increases the threshold input power for SP. 
For $\alpha^\prime\le 0.93$ (approximately), we do not observe any SP for a given interval of input powers. 

\begin{figure}[ht]
\includegraphics[width=7.7cm]{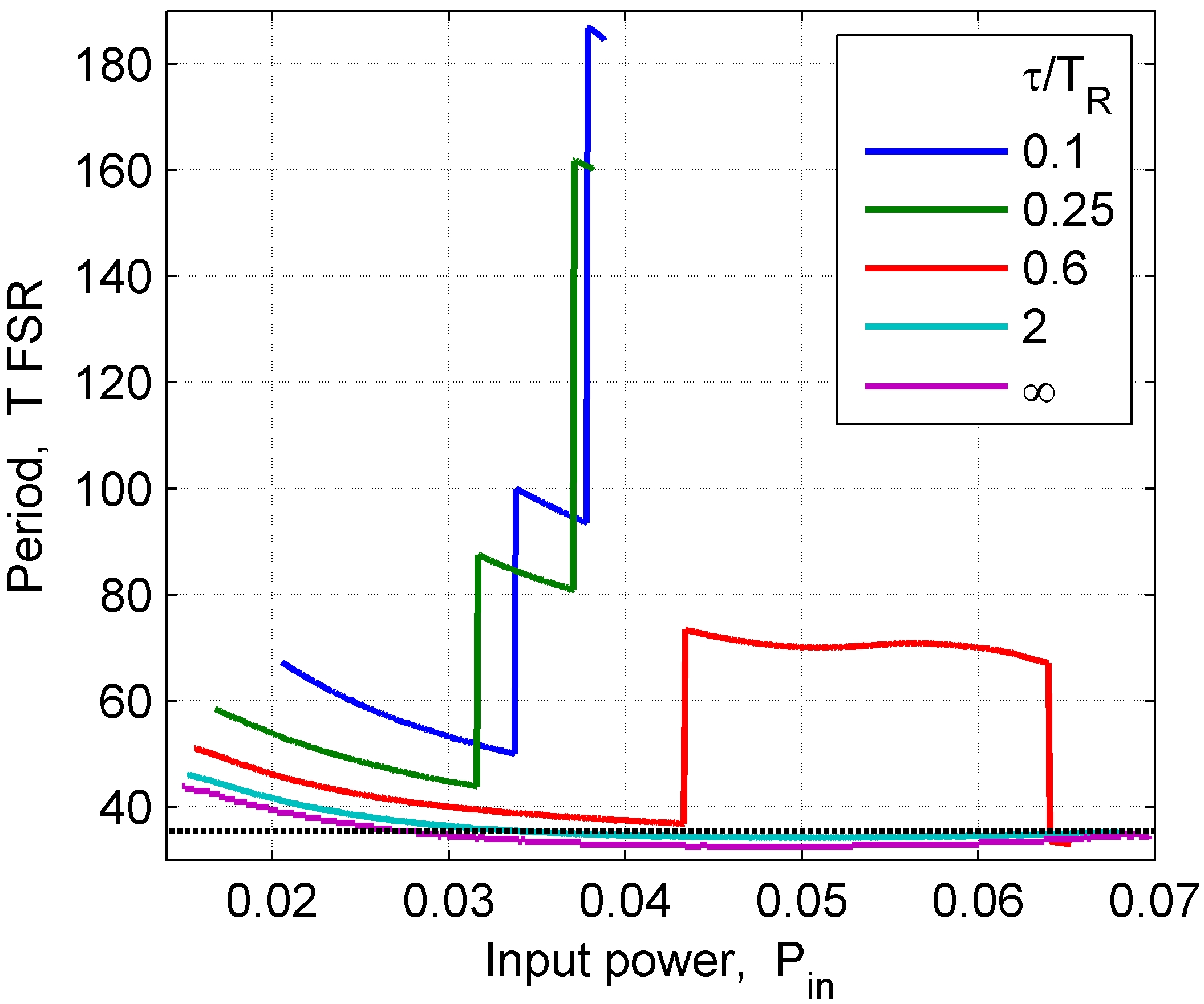}
\caption{Normalized period $T\,\rm{FSR}$ of SP solution vs. input power $P_{\rm{in}}$ for the various values of $\tau/T_{\rm{R}}$ shown in the legend; other structural parameters are as in Fig. 3.
The black dashed line corresponds to the beating frequency of the two linear modes.}
\label{fig:responsetime}
\end{figure}

In Fig. \ref{fig:responsetime}, we demonstrate the influence of the input power on the SP period. This example also demonstrates the effect of finite relaxation time $T_{\rm{R}}$ on the SP behavior. 
For small values of $\tau/T_{\rm{R}}$, we observe period-doubling followed by chaos. The SP region increases with $\tau/T_{\rm{R}}$, mainly due to the shift of the upper limit of $P_{\rm{in}}$. (Chaotic states, that occur above this limit, are not shown in Fig. \ref{fig:responsetime}.) With a further increase of $\tau/T_{\rm{R}}$, the period-doubling vanishes, see the curve for $\tau/T_{\rm{R}}=2$, and the dependence approaches the results obtained in the approximation of instantaneous response, $\tau/T_{\rm{R}}\rightarrow\infty$. However, this should be taken with caution; the approximation of instantaneous response can break down for longer time scales \cite{Ikeda2}. As a result, more complicated SP behavior with period $T\,\rm{FSR}\approx 1$, (not shown in Fig. \ref{fig:responsetime}) is revealed when $\tau/T_{\rm{R}}$ is increased above about $3$. 

Finally, we compare our results with the phenomenological model that considers beating of the linear modes in coupled cavities \cite{grigoriev}. 
It follows from Ref. \cite{grigoriev} that the period of SP can be estimated by using the formula $T=2/(f_1-f_2)$, where $f_1$ and $f_2$ are frequencies of the linear modes in a two-cavity system. For our system, these frequencies correspond to the two peaks seen in Fig. \ref{fig:map}(a) in the limit $P_{\rm{d}}\rightarrow 0$. The calculated period, which is shown in 
Fig. \ref{fig:responsetime} (horizontal dashed line), qualitatively matches our exact results for the regime in which the approximation of instantaneous response can be applied.

\section{Conclusions}
\label{conc}
We presented the difference-differential equations that describe the time evolution in Kerr-nonlinear resonant structures consisting of coupled ring resonators. The model includes both 
the effect of loss and non-instantaneous Kerr response. We formulated a simple and efficient numerical technique for the solution of the system.

To demonstrate the technique, we investigated the dynamical behavior of a simple structure with two rings. We presented steady-state solutions and their classification. We identified suitable parameters for self-pulsing operation and demonstrated the self-pulsing state and the corresponding limit cycle. 

Both the  loss  and/or the finite relaxation time can significantly affect the range of input powers for observing self-pulsing states. In addition, the self-pulsing period strongly changes with the finite relaxation time. In some circumstances, however, the period can be estimated by considering the beating of linear modes in coupled cavities \cite{grigoriev}. 

\section*{Acknowledgments}
This work was supported by The Grant Agency of the Academy of Sciences of the Czech Republic under contract IAA101730801. 
Y. Ek\c{s}io\u{g}lu acknowledges the support by the project CZ.1.07/2.3.00/30.0039 of Brno University of Technology.

\end{document}